\documentclass[letterpaper, 10pt, conference]{ieeeconf}
\IEEEoverridecommandlockouts 
\usepackage{amssymb}
\usepackage{amsmath}

\usepackage{amsthm}
\usepackage{cite}
\usepackage{graphics}
\usepackage{hyperref}
\usepackage{multirow}
\usepackage{tikz}
\usepackage{graphicx}

\newtheoremstyle{named}{}{}{}{}{\bfseries}{.}{.5em}{\thmnote{#3} Principle}
\theoremstyle{named}

\theoremstyle{definition}

\def\u{\mathbf{u}}
\def\w{\mathbf{w}}
\def\x{\mathbf{x}}
\def\K{\mathbf{K}}
\def\x{\mathbf{x}}
\def\xhat{\hat{\mathbf{x}}}
\def\deltahat{\hat{\boldsymbol{\delta}}}

\def\R{\mathbf{R}}
\def\M{\mathbf{M}}

\title{\LARGE \bf
Internal Feedback in Biological Control: \\
Locality and System Level Synthesis
}

\author{Jing Shuang (Lisa) Li
	\thanks{J. S. Li is with Computing and Mathematical Sciences, California Institute of Technology. {\tt\small jsli@caltech.edu}. This research was in part supported by NSERC PGSD3-557385-2021.
	}%
	\thanks{This paper is one of three in a series on internal feedback in biological control architectures. These papers may be read in any order, though a suggested order is \cite{Paper1}, \cite{Paper2}, then this paper.}
}
{\tiny }
\begin{document}	
	\maketitle
	\thispagestyle{empty}
	\pagestyle{empty}
	
	\begin{abstract}

The presence of internal feedback pathways (IFPs) is a prevalent yet unexplained phenomenon in the brain. Motivated by experimental observations on \textit{1)} motor-related signals in visual areas, and \textit{2)} massively distributed processing in the brain, we approach this problem from a sensorimotor standpoint and make use of distributed optimal controllers to explain IFPs. We use the System Level Synthesis (SLS) controller to model neural phenomena such as signaling delay, local processing, and local reaction. Based on the SLS controller, we make qualitative predictions about IFPs that strongly align with existing experimental observations. We introduce a `mesocircuit' for optimal performance with distributed and local processing, and local disturbance rejection; this mesocircuit requires extreme amounts of IFPs and memory for proper function. This is the first theory that replicates the massive amounts of IFPs in the brain purely from \textit{a priori} principles, providing a new theoretical basis upon which we can build to better understand the inner workings of the brain.

\end{abstract}
	\section{Introduction} \label{sec:introduction}

\begin{figure}
	\centering
	\includegraphics[width=7cm]{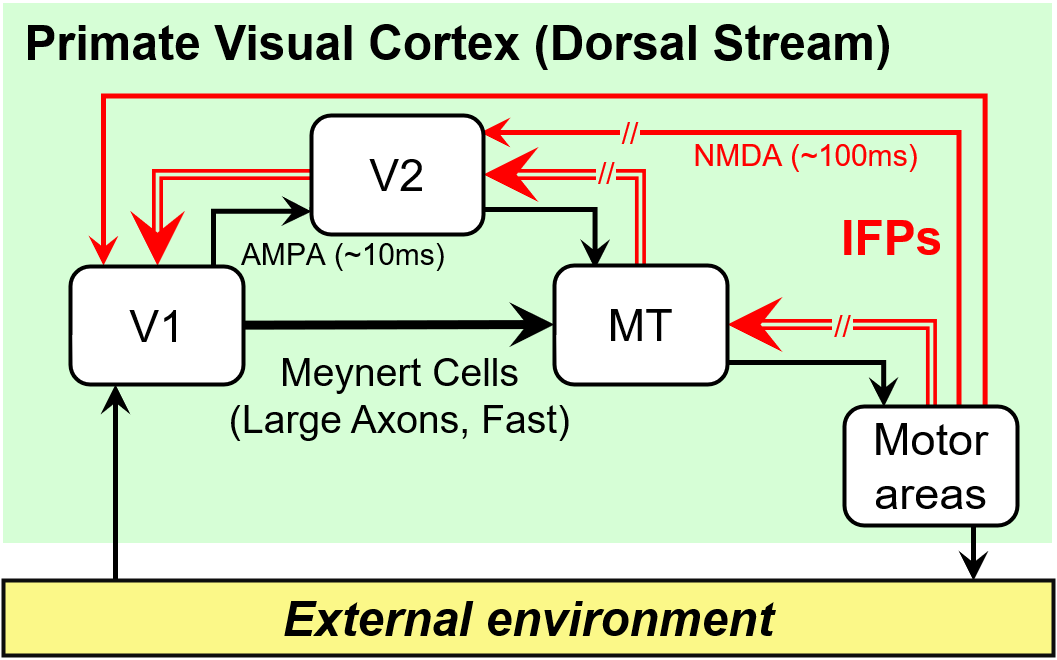}
	\caption{Primate visual cortex contains large amounts of IFPs, which vastly outnumber forward pathways \cite{Budd1998}. Forward pathways are shown in black; internal feedback pathways (IFPs) are shown in red. The forward direction travels from sensing toward actuation (left to right); internal feedback travels in the reverse direction. Compound-line arrows indicate larger signals/multiple connections (e.g. vectors instead of scalars). Not shown is the massively distributed processing that also occurs in primate visual cortex and everywhere in the brain \cite{Felleman1991}.}
	\label{fig:sensorimotor_feedback}
	\vspace{-1.5em}
\end{figure}

The primate visual pathway propagates visual input from the eye to the brain. Information travels from the retina in the eye to the lateral geniculate nucleus (LGN), then to the primary visual area (V1) in the cortex, secondary visual area (V2), and so on. However, massive amounts of connections in the reverse direction (i.e. internal feedback pathways, or IFPs) are also observed, as shown in Fig. \ref{fig:sensorimotor_feedback}. These connections are known by a variety of names in neuroscientific literature, including \textit{descending feedback}, \textit{predictive feedback}, \textit{reciprocal connections}, and \textit{recurrence}. Feedback is well-documented but poorly understood \cite{Felleman1991, Callaway2004, Muckli2013}; understanding the purpose of this mechanism is invaluable to understanding overall circuit function in the visual system. 

A substantial portion of neural activity in the visual areas are unrelated to visual input, but instead dominated by body movements \cite{Stringer2019, Musall2019}. Motivated by these recent findings, we propose an analysis of IFPs from a sensorimotor perspective. Optimal control models are an ideal candidate for this analysis as they are widely successful in explaining behavior-level observations in the sensorimotor domain \cite{Todorov2004, Franklin2011}. A desirable next step is to use optimal control to unravel the mysteries of the \textit{architecture} and structural function of the sensorimotor system. However, this presents a major challenge: biological controllers use cells and neurons, which are much more constrained than the fast electronic components used in man-made controllers. Some key differences are listed in Table \ref{table:ideal_vs_constrained}. To accommodate these constrained components, new theory and analysis is required.

\begin{table}[h]
	\caption{Ideal vs. constrained controller components \vspace{-1em}}
	\label{table:ideal_vs_constrained}
	\begin{center}
		\begin{tabular}{|l|l|l|}
			\hline
			& \textbf{Communication} & \textbf{Actuation \& Sensing} \\
			\hline \hline
			\textbf{Ideal} & dense, fast, global & dense, fast \\
			\hline
			\textbf{Constrained} & sparse, delayed, local & sparse, delayed \\
			\hline
		\end{tabular}
	\end{center}
\vspace{-1.5em}
\end{table}

We incorporate constraints described in Table \ref{table:ideal_vs_constrained} into optimal control theory, and study the IFPs in the resulting controllers. In our companion paper \cite{Paper2}, we model basic actuation and sensing delay using a variation of standard optimal control, and demonstrate that IFPs are required for optimal function when such delays are present. However, standard theory cannot capture local, distributed, and delayed communication \textit{within the controller} -- this requires the newer System Level Synthesis (SLS) theory \cite{Anderson2019}. In this paper, we use an SLS controller to capture local, distributed, and delayed communication, and show that it contains extreme amounts of IFPs that match neural anatomy \cite{Budd1998}. In Section \ref{sec:sparsity}, we discuss the effect of sparse closed-loops and sparse controllers, and use an intermediate parametrization to introduce the notion of local memory patches. We then leverage these concepts in Section \ref{sec:mesocircuit} to describe the full SLS-based mesocircuit, and examine the massive amounts of IFPs that are intrinsic to this mesocircuit. Section \ref{sec:conclusion} contains closing comments. We remark that our goal is to provide standalone theory, which can highlight fruitful experimental directions and guide interpretations of large-scale data. Though we do not include empirical data in this work, we discuss possible experiments and biological systems of interest in \cite{Paper1}.
 
IFPs are not only present in the visual system; they are prevalent in neural systems (e.g. somatosensory and motor cortices \cite{Felleman1991}; auditory system \cite{Suga2008}) and cellular systems (\cite{Paper1} and references therein). Though we focus on the sensorimotor system, our analysis is sufficiently general to be useful for other systems. Additionally, understanding the function of IFPs is beneficial for the design of modern cyberphysical systems. Although fast electronic parts traditionally allowed engineers to avoid the constraints in Table \ref{table:ideal_vs_constrained}, such constraints become unavoidable as we move toward efficient, sustainable, large networked systems. Thus, understanding the role of IFPs in architectures with constrained components allows us to design efficient, sustainable, large-scale controllers.

	\section{Closed Loops, Sparsity, and Memory} \label{sec:sparsity}

We focus on the state feedback problem with local and distributed processing in the controller, and local reactions in the closed-loop. Though many different mechanisms (e.g. delay, internal estimation, as per \cite{Paper2}) necessitate IFPs, local and distributed processing necessitates the largest amount of IFPs. By observing mechanisms one at a time, we can produce human-interpretable models that we will eventually combine to create more complete (albeit indecipherably complex, like the brain) and quantitative models. 

\subsection{Closed Loops and Sparsity}
Consider the linear time-invariant (LTI) discrete-time system described in frequency domain by
\begin{equation}
	z\x = A\x + B\u + \w
\end{equation}
\noindent where $\x$, $\u$, and $\w$ denote state, input, and disturbance, respectively. We sense the state and apply a linear transfer matrix $\K$ as the controller, i.e. $\u=\K\x$, and define the resulting closed-loop responses $\R, \M$ as in \cite{Anderson2019}:
\begin{equation} \label{eq:closed_loop_responses}
\begin{bmatrix} \x \\ \u \end{bmatrix} =
\begin{bmatrix} (zI - A - B\K)^{-1} \\ \K(zI - A - B\K)^{-1} \end{bmatrix} \mathbf{w} =:
\begin{bmatrix} \R \\ \M \end{bmatrix} \mathbf{w}
\end{equation}
$\R, \M$ are transfer matrices, and can be written as a sum of spectral elements (constant matrices) $R(k), M(k)$:
\begin{equation} \label{eq:spectral_components}
	\R = \sum_{k=0}^{\infty} R(k)z^{-k}, \quad \M = \sum_{k=0}^{\infty} M(k)z^{-k}
\end{equation}

We can also write $\K$ in terms of spectral elements, though $\K$ is often static; for LQR, we solve for constant matrix $K_{LQR}$ and set $K(0) = K_{LQR}$, $K(k) = 0$ for $k > 0$. 

Spectral elements of $\R, \M$ directly relate to responses to impulse disturbances. Suppose the system is perturbed by an impulse at node $i$, time $t$. Then, the actuation impulse response from time $t$ to $t+k$, i.e. $u(t:t+k)$, can be obtained by looking at the $i$th column of $M(0), M(1) \dots M(k)$. For simplicity, we will work with finite-horizon $\R, \M$, i.e. $M(k) \equiv 0$ for $k > T$ for some finite horizon length $T$.

Typically, we design and implement the controller $\K$. We want $\K$, in conjunction with the plant, to give a closed-loop system that performs optimally. In large systems, we also want some sparsity in the controller $\K$ and closed-loop responses $\R$ and $\M$. We focus on local communication and local reaction as the two main sources of sparsity constraints; both are abundant in biological systems. 

\subsubsection{Local communication}
Neurons communicate locally; each neuron communicates with a handful of nearby neurons, which comprise a very tiny subset of the organism's total neurons. Local communication imposes local sparsity structure on the controller (e.g. $\K$). Local communication is desirable in large cyberphysical systems such as the smart grid, in which centralized communication is unwieldy.

\subsubsection{Local disturbance rejection}
Biological systems react locally; reflexes contain and reject disturbances using only local actuators, without affecting the entire organism. Local reaction imposes local sparsity structure on the closed-loop responses $\R$ and $\M$. Local reaction is desirable in systems such as the smart grid, where we do not want a disturbance at one part of the grid to spread to the entire grid.

In an ideal system (see Table \ref{table:ideal_vs_constrained}), we obtain sparsity for free, i.e. without specifying it as a constraint or objective. For an LQR problem with no input penalty and full actuation (i.e. $B=I$), the optimal controller $K$ has the same sparsity as the state matrix $A$, and gives rise to similarly sparse deadbeat closed-loop responses $\R, \M$. However, in a non-ideal case (e.g. sparse actuation instead of full actuation), we no longer obtain this desired sparsity. To demonstrate, we use a symmetric 8-node ring, shown in the top left of Fig. \ref{fig:local_m_memory}. The state matrix $A \in \mathbb{R}^{8 \times 8}$ and actuation matrix  $B \in \mathbb{R}^{8 \times 4}$ are:
\begin{equation} \label{eq:system_matrices}
A = \frac{a}{3} *
\begin{bmatrix} 
1 & 1 & 0 & \dots & 1\\
1 & 1 &  1 & 0 &  \dots  \\
0  & \ddots & \ddots &  \ddots &  \vdots \\
\vdots & \ddots & \ddots & \ddots & 1\\
1 &     0   & \dots & 1 &  1
\end{bmatrix}, B = \begin{bmatrix} 
1 & 0 & 0 & 0 \\
0 & 0 & 0 & 0 \\
0 & 1 & 0 & 0 \\
& \ldots & \ldots & \\
0 & 0 & 0 & 0 
\end{bmatrix}
\end{equation}

\noindent where the spectral radius of $A$ is equal to $a$; we set $a=1.8$ to obtain an unstable plant. 50\% of nodes are actuated; nodes 1, 3, 5, 7 receive actuation, while nodes 2, 4, 6, 8 do not.

Consider an LQR problem with state penalty $Q=I$ and negligible input penalty\footnote{i.e. $R = \epsilon I$ for some $\epsilon \ll 1$. We omit input penalty from our analysis in order to simplify the problem presentation; all results in this paper hold for arbitrary input penalty.}:

\begin{equation} \label{eq:lqr_k}
\begin{aligned}
\min_{K} \quad & \sum_{t=0}^{\infty} x(t)^{\top}Qx(t) \\
\textrm{s.t.} \quad &  x(0) = x_0, \quad u(t) = Kx(t) \\
& x(t+1) = Ax(t) + Bu(t) + w(t) \\
\end{aligned}
\end{equation}

\begin{figure}
	\centering
	\includegraphics[width=8cm]{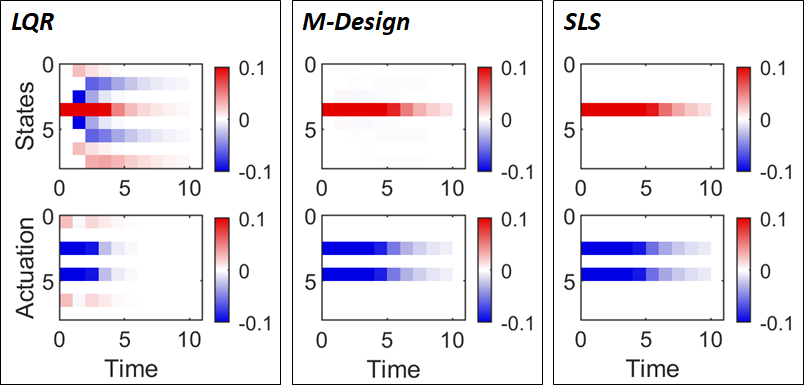}
	\caption{Impulse responses for the 8-node ring system using the LQR, M-Design, and SLS controllers. At $t=0$, an impulse disturbance hits node 4. The LQR controller gives a global impulse response; every state and actuator reacts to this disturbance. The M-Design and SLS controllers give local impulse responses (i.e. local closed-loop responses); only nodes 3, 4, and 5 react to this disturbance. Associated normalized costs: 1.000 for LQR, 1.035 for M-Design, and 1.037 for SLS. Despite being highly localized, the M-Design and SLS controllers perform nearly optimally.}
	\label{fig:impulse_responses}
	\vspace{-1.5em}
\end{figure}

Optimal $K$ is obtained by solving the DARE. In our example, neither the controller nor closed-loop responses are sparse, as shown in the ``LQR'' panels in Fig. \ref{fig:controller_sparsity} and Fig. \ref{fig:impulse_responses}). Worse, we are unable to specify any sparsity constraints on the closed-loop responses $\R, \M$, since such constraints would be nonconvex in $K$ as per \eqref{eq:closed_loop_responses}. The problem of specifying sparsity on the closed-loop responses in a convex manner is addressed by the SLS parametrization, which we will use in the next section. However, for tutorial purposes, and to introduce the presence of \textit{memory} in our eventual mesocircuit, we will first consider an intermediate parametrization which we call \textit{M-Design}.

\subsection{M-Design}
We want to design a controller and closed-loop responses $\R, \M$ such that all are sparse. Unfortunately, sparsity constraints on $\R, \M$ are nonconvex in $K$, as described above.

Assume that instead of sensing the state $\x$, we sense the disturbance $\w$. Then, we can apply a disturbance feedback controller $\u=\M\w$. We emphasize that this is merely a tutorial step; we do not generally expect disturbance sensing to be possible, and we do not use this controller in practice.

Under disturbance feedback, $\M$ is both the controller \textit{and} the closed-loop actuation response. We can now reformulate the LQR problem in Eq. \ref{eq:lqr_k}, with $\M$ as the optimization variable instead of $K$; this is the \textit{M-Design} problem:

\begin{equation} \label{eq:m_design}
\begin{aligned}
\min_{\R, \M} \quad &  \| Q^{\frac{1}{2}}\R\|^2 \\
\textrm{s.t.} \quad & \begin{bmatrix} zI-A & -B \end{bmatrix} \begin{bmatrix} \R \\ \M \end{bmatrix} = I \\
 & \R, \M \text{ stable, causal}, \quad \R, \M \in \mathcal{S}
\end{aligned}
\end{equation}

The first constraint in \eqref{eq:m_design} is the \textit{feasibility constraint}; it ensures that the proposed controller and closed-loop responses obey the original system dynamics. In practice, we solve \eqref{eq:m_design} by optimizing over spectral elements of finite-horizon $\R, \M$; the feasibility constraint translates to $R(k+1) = AR(k) + BM(k)$. We can additionally constrain $\R, \M$ to be sparse via the set $\mathcal{S}$; this results in sparsity constraints on spectral elements. Note that in M-Design, these sparsity constraints are affine. Since $\M$ is both the controller and closed-loop actuation response, $\M \in \mathcal{S}$ gives sparsity in both the controller and one of the closed-loops; $\R \in \mathcal{S}$ gives sparsity in the other closed-loop. In contrast, we could not enforce sparsity on any closed-loop response in the original LQR problem \eqref{eq:lqr_k}, as it resulted in nonconvexity. 

\begin{figure}
	\centering
	\includegraphics[width=6cm]{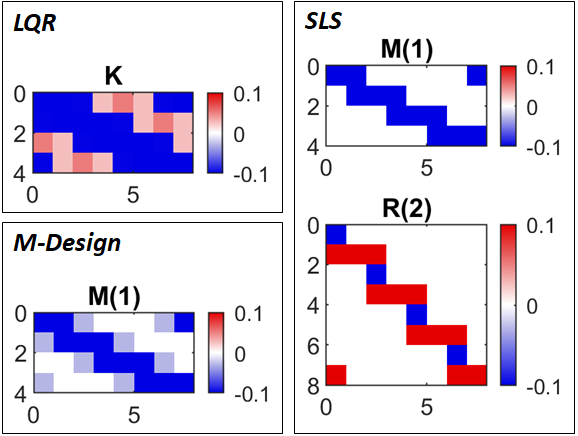}
	\caption{Sparsity of the LQR controller, M-Design controller, and SLS controller, for the 8-node ring system. All spectral elements $M(k)$ are constrained to have the same sparsity, so we show $M(1)$ as a representative element. For $\R$, all spectral elements are constrained to have the same sparsity except $R(1)$, which the SLS feasibility constraint restricts to be the identity matrix; we show $R(2)$. The LQR controller is dense, while the other two are constrained to be local. We remark that the SLS controller $\M$ appears to be sparser than the $\M$ matrix from M-Design, even though we applied the same sparsity constraint to both.}
	\label{fig:controller_sparsity}
	\vspace{-1.5em}
\end{figure}

We now use M-Design to specify 2-hop local sparsity; each node may only communicate with its neighbors and its neighbors' neighbors. For the ring system, this means each node communicates with two neighbors on each side (e.g. Node 4 communicates with Node 2, 3, 5, 6). The resulting controller and closed-loops are shown in the ``M-Design'' panels in Fig. \ref{fig:controller_sparsity} and Fig. \ref{fig:impulse_responses}. We see that the desired sparsity is indeed achieved. Furthermore, the performance of this sparse controller is only 4\% worse than that of the optimal controller, despite the drastic difference in sparsity.

As shown above, M-Design accommodates sparsity motivated by local communication. M-Design also accommodates sparsity motivated by delayed communication, which is a common feature of biological systems; communication via neurons is orders of magnitude slower than communication via electronics, and slower neurons are often preferred because they are less expensive to maintain \cite{Sterling2015}. For large cyberphysical systems, delay-tolerant controllers are important as the distances spanned by such systems make instant global communication impossible.

\subsection{Local memory patches}
We introduce the notion of a \textit{local memory patch} using the M-Design controller. Though the M-Design controller is completely comprised of forward pathways, the local memory mechanism will play an important role in the SLS mesocircuit with IFPs in the next section.

First, we write $\u=\M\w$ as a time-domain convolution. We assume $\M$ to have some finite horizon length $T$:
\begin{equation}
u(t) = \sum_{k=0}^{T} M(k)w(t-k)
\end{equation}

\begin{figure}
	\centering
	\includegraphics[width=8.5cm]{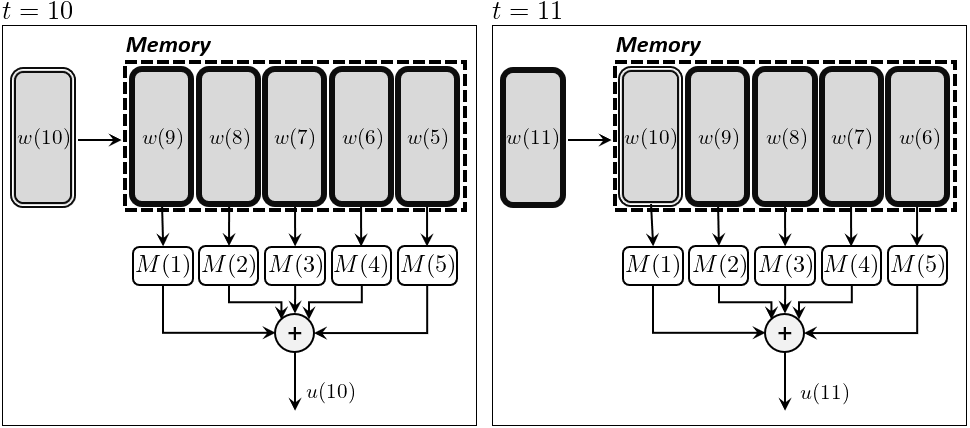}
	\caption{Standard implementation of a system with FIR transfer function $\M$, input $\w$, and output $\u$. This example uses a strictly causal $\M$ with horizon length $T=5$. \textit{(Left)} At time $t=10$, $w(t)$ for $t=5:9$ are in memory. The output $u(10)$ is produced by summing $M(k)w(t-k)$ for $k=1:5$. \textit{(Right)} At the next time step, $t=11$, the entries in memory shift to the right. The oldest value, $w(5)$, is discarded (i.e. forgotten), and the newest value, $w(10)$, enters the memory from the left. The output $u(11)$ is again produced by summing products of $M(k)w(t-k)$.}
	\label{fig:memory_shifting}
	\vspace{-1.5em}
\end{figure}

To compute $u(t)$, we require a memory of the most recent $T$ values of $w$. Every time step, we discard the oldest value of $w$ from memory and add the newest value of $w$. An example of this is shown in Fig. \ref{fig:memory_shifting}. This is a standard implementation of a system with FIR transfer matrix $\M$; we will build upon it to create locally implemented controllers and memory patches. We remark that although increased $T$ generally improves performance, it requires more memory. 

Suppose we solve \eqref{eq:m_design} with local sparsity constraints on $\M$. We can implement $\M$ using some standard realization. This gives sparse closed-loop responses, as desired, but may not preserve communication sparsity; the resulting controller may communicate globally instead of locally, which we don't want. We now provide a natural realization that preserves sparse communication, and defer exploration of other sparsity-preserving controller realizations to future work.

\begin{figure}
	\centering
	\includegraphics[width=8.5cm]{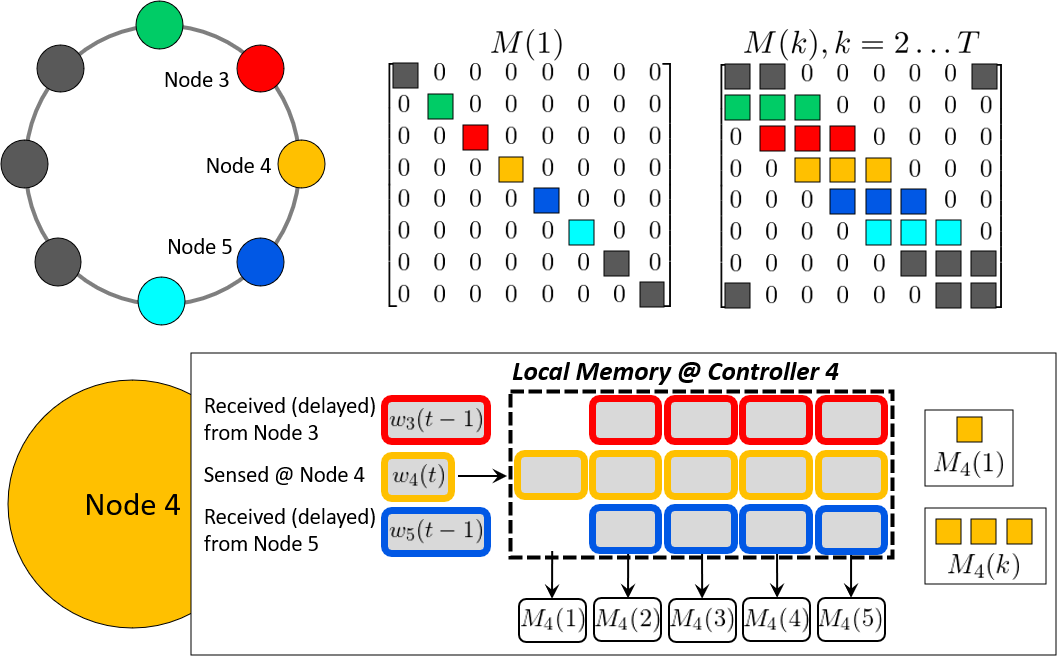}
	\caption{Note: for ease of visualization, we assume all nodes are actuated. \textit{(Top left)} Ring with 8 nodes. \textit{(Top right)} Spectral elements of $\M$. Colored squares represent nonzero values; all other values are constrained to be zero. Nonzero values away from the diagonal represent communication between nodes. Sparsity constraints arise from delayed communication (for $M(1)$) and local communication (for $M(k), k>2$). Sparsity on $\M$ additionally translates to local disturbance rejection. \textit{(Bottom)}. Local controller and memory patch at node 4. Each node uses its own row of $M(k)$ to implement its local controller. Rectangles in local memory represent scalar values of $w_i(t)$; colors indicate the source of the value, e.g. red rectangles are $w_3$ values from node 3. Recent entries are toward the left, and oldest entries are toward the right. Local actuation (not shown) is produced by multiplying $M_4(k)$ by columns in memory and summing over the products.}
	\label{fig:local_m_memory}
\end{figure}

We implement $\M$ distributedly: each node implements its own controller with its own local memory patch. Each node $i$ only senses disturbance $w_i$, and relies on local inter-node communication to access $w_j$ for $j \neq i$. We show an example in Fig. \ref{fig:local_m_memory}, where each node communicates only with its immediate neighbors. In our example, we consider inter-node delays, but not self delays (i.e. sensor $i$ communicates to controller $i$ with some nonzero delay). Self delays can be easily enforced via zero-constraints on the diagonals of $M(k)$ for appropriate $k$; in this way, M-Design (and later, SLS) can accommodate all delays that can be modeled using standard control theory, including those in \cite{Paper2}.

Since each node has its own local memory, a lot of redundant memory is created. In our example, the memory at node 4 stores past values of $w_3$; past values of $w_3$ are also stored at node 3 and node 2. Here, each disturbance $w_i$ is stored 3 times: by node $i$ and its neighbors $i+1$ and $i-1$. For a general system, each disturbance $w_i$ will be stored by node $i$ and all neighbors with whom it communicates.

\begin{figure}
	\centering
	\includegraphics[width=7.5cm]{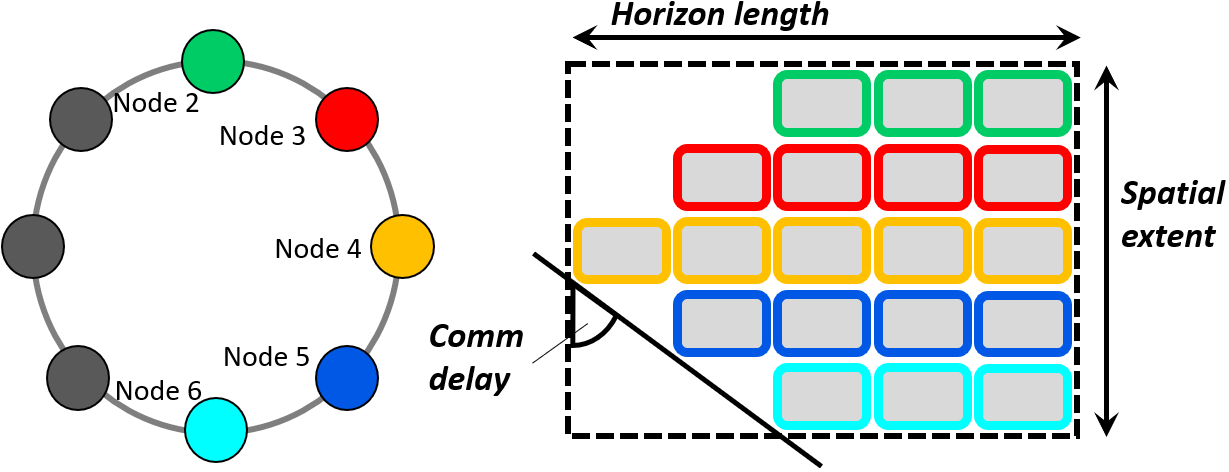}
	\caption{\textit{(Left)} Ring with 8 nodes. \textit{(Right)} Size and shape of example local memory patch at node 4. Recent information is on the left, old information is on the right. Horizon length $T$ indicates how far to remember into the past. Spatial extent indicates how many neighbors each node communicates with. For the ring system, communication delay is indicated by the angle of the triangular `front' of the memory; larger angle corresponds to larger delay. In this example, $T=5$, and nodes communicate to their 4 nearest neighbors with delay proportional to distance. Node 4 has up-to-date information on node 4 (yellow), slightly outdated (delayed one time step) information from nodes 3 and 5 (red and blue), and more outdated (delayed 2 time steps) information from nodes 2 and 6, which are farther away.}
	\label{fig:memory_dimension}
    \vspace{-1.5em}
\end{figure}

Fig. \ref{fig:local_m_memory} shows an example of a local memory patch at a single node. A more general characterization is shown in Fig. \ref{fig:memory_dimension}. In our example, the size of the local memory patches are uniform across nodes; however, in general, the size of the local memory patches may vary from node to node. All of this is supported by the M-Design formulation; one must only specify the appropriate sparsity constraint on $\M$ in \eqref{eq:m_design}.

Any controller that is expressed as a transfer function can be realized and implemented in a variety of ways. Our case is unique, as we impose implementation-level (e.g. communication) constraints on the controller; this is generally incompatible with standard realization theory. Nonetheless, alternative implementations for our controller exists. For instance, instead of an explicit local memory patch, memory may be implicitly contained in delayed wires; instead of node 4 having $w_3(t)$, $w_3(t-1)$, and $w_3(t-2)$ in a local memory patch, there can simply be three wires from node 3 to node 4 with delays of 0, 1, and 2, and so on. The question of general system realization and implementation with implementation-level constraints is deferred to future work.

	\section{Mesocircuits for Internal Feedback} \label{sec:mesocircuit}

We use SLS to produce a mesocircuit that predicts large amounts of IFPs relative to forward pathways, purely from \textit{a priori} principles. This prediction is consistent with observations from neuroscience \cite{Callaway2004, Muckli2013, Budd1998} and also incorporates local communication, local disturbance rejection, and signaling delay; as previously described, these are ubiquitous features of the nervous system. We make extensive use of the notions of locality and memory from the M-Design controller.

A \textit{\textbf{mesocircuit}} refers to a physiological circuit in an organism. It consists of $\sim$thousands (or more) of cells working together to provide some functionality. A mesocircuit is a larger-scale entity than a \textit{microcircuit}, which concerns individual cells and neurons; but a smaller-scale entity than the circuit shown in Fig. \ref{fig:sensorimotor_feedback}, in which circuit components (e.g. V1) are composed of $\sim10^7$ neurons \cite{Colonnier1981}. In our case, a single `node' in our analysis may correspond to a cluster of cells or neurons in an organism.

\subsection{System Level Synthesis}
In M-Design, we assume disturbance-sensing capabilities, and implement a disturbance feedback controller. This allows us to specify sparsity constraints on controller and closed-loop responses in an affine manner. Now, we remove the disturbance-sensing assumption and revert to state feedback. As before, we want to impose sparsity constraints on the controller and closed-loop responses: we do this using System Level Synthesis (SLS) \cite{Anderson2019}. The SLS reformulation of the original LQR problem \eqref{eq:lqr_k} is nearly identical to the M-Design problem \eqref{eq:m_design}, with one difference -- $\R, \M$ must now be \textit{strictly} causal instead of causal, i.e. $R(0) \equiv 0, M(0) \equiv 0$. This is because in disturbance feedback, $w(\tau)$ may affect $x(t)$ and $u(t)$ casually, i.e. for $t \geq \tau$; but in state feedback, the relationship is strictly causal, i.e. $t > \tau$.

\begin{figure}
\begin{minipage}{4cm}
	\includegraphics[width=3.5cm]{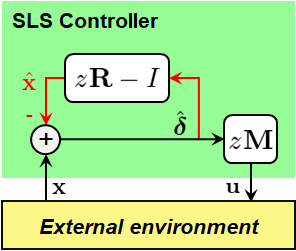}
\end{minipage}
\begin{minipage}{4.5cm}
		\caption{Block diagram for the state feedback SLS controller. Sparsity in $\R$ and $\M$ directly translate to sparsity in the controller. The IFP (in red) takes the disturbance estimate $\deltahat$ and produces state estimate $\xhat$; this is subtracted from the sensed state $\x$ to produce the disturbance estimate. All blocks are causal.} \label{fig:sls_basic}
\end{minipage}
\vspace{-1.5em}
\end{figure}

Though the optimization for M-Design and SLS are nearly identical, the implementation is drastically different. In M-Design, we directly use $\u = \M\w$; in SLS, we use both $\R$ and $\M$ to implement the controller, as shown in Fig. \ref{fig:sls_basic}. Feasibility, stability, and robustness of this formulation and implementation are discussed at length in \cite{Anderson2019}.

SLS is a natural next step to M-Design. Instead of sensing the disturbance, we use an IFP (see Fig. \ref{fig:sls_basic}) to produce an estimate of it. The IFP signal is $\xhat$, the predicted future state; this coincides with prevailing notions of IFPs in neuroscience \cite{Muckli2013, Huang2011}. We subtract this predicted state from the sensed state to calculate the disturbance estimate $\deltahat$. We then use $\deltahat$ to generate the control output via the $z\M$ block; it is also used to generate the predicted future state via the $z\R-I$ block. Since we are in state feedback, the `estimations' are exact; $\hat{\delta}(t) = w(t-1)$, and $\hat{x}(t+1) = x(t+1)$ if $w(t)=0$. Also, the prediction done by the IFP implicitly uses knowledge of past sensed states and past actions.

In M-Design, we use $\M$ as both the controller and closed-loop response; in SLS, both $\R$ and $\M$ are controller and closed-loop response. The feedback structure in Fig. \ref{fig:sls_basic} ensures that sparsity in $\R, \M$ translates directly to sparsity in the controller. Additionally, by Thm 4.1 in \cite{Anderson2019}, the closed-loop responses of the system are also $\R$ and $\M$. As before, we can enforce sparsity constraints on $\R, \M$ via $\mathcal{S}$; these constraints are affine. We use SLS to design a controller for the 8-node ring that is sparse (localized) in both controller and closed-loop response, shown in the ``SLS'' panels of \ref{fig:controller_sparsity} and \ref{fig:impulse_responses}. The performance of this sparse controller is less than 4\% worse than that of the optimal, non-sparse controller.

We remark that in this SLS formulation, any constraint on $\R, \M$ will constrain both the controller and the closed-loop. In our example, we use this to our advantage; however, there are cases where we don't want to constrain both. In such cases, we may use alternate formulations, such as two-step SLS \cite{Li2020} or virtually localized SLS \cite{Matni2017}. These methods utilize the same controller structure as standard SLS, shown in Fig. \ref{fig:sls_basic}. Thus, the resulting controller structure remains unchanged, and the discussion on memory, mesocircuits, etc. still apply to these alternative methods.

\begin{figure}
	\centering
	\includegraphics[width=8.5cm]{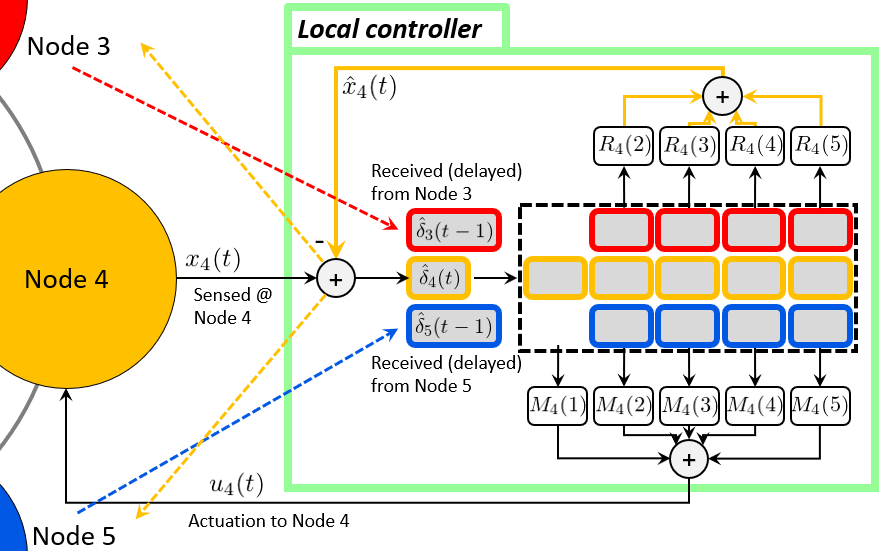}
	\caption{Local SLS controller at node 4, assuming node 4 is actuated. Memory is used by both the forward path (through $\M$) and predictive IFP (through $\R$). Communicative IFPs to and from node 4 are shown by dashed lines. We enforce inter-node communication delay; $\hat{\delta}_3$ and $\hat{\delta}_5$ are received from neighboring nodes with a delay of 1 time step. Note: $R(1)$ is not included because for all $\R$ satisfying the feasibility constraint, the IFP expression $z\R-I$ (see Fig. \ref{fig:sls_basic}) results in $R(1)$ being canceled out.}
	\label{fig:single_microcircuit}
	\vspace{-1.5em}
\end{figure}

\subsection{Full Mesocircuit}

The full mesocircuit at a single node is shown in Fig. \ref{fig:single_microcircuit}. Here, elements in memory represent estimated disturbances $\hat{\delta}$. The forward path passes from sensing input, through the memory patch, through $M$, toward the actuation output. One IFP is the connection through $R$, from the memory patch toward the sensory input; we refer to this as predictive IFP. Inter-node communication, shown by the dashed arrows, can also be interpreted as IFPs; we refer to these as communicative IFPs. Communicative IFPs are analogous to \textit{lateral connections} or \textit{lateral feedback} in neuroscience, which are generally classified as IFPs in neuroscience.

\begin{figure}
\begin{minipage}{5.1cm}			  
	\includegraphics[width=5cm]{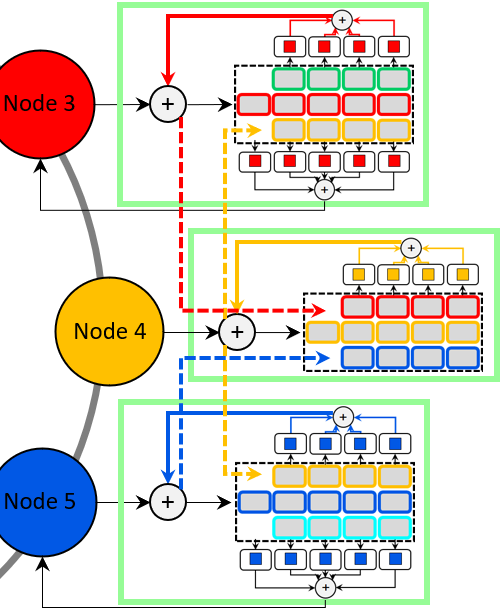}
\end{minipage}
\begin{minipage}{3cm}
	\caption{Local SLS controllers at nodes 3, 4, and 5. Each local controller is enclosed in a light green box. Only nodes 3 and 5 are actuated. This figure contains 2 forward paths (one per actuated node), 3 predictive IFPs (one per node), and 4 communicative IFPs, shown by dashed lines. Not shown are communicative IFPs to and from other neighbors of nodes 3 and 5.}
	\label{fig:multiple_microcircuit}
\end{minipage}
	\vspace{-1.5em}
\end{figure}

We show three local SLS mesocircuits in Fig. \ref{fig:multiple_microcircuit}. Here, node 4 is unactuated; however, despite having no actuation, node 4 still requires memory and circuitry to calculate $\hat{\delta}_4$ and communicate it to its neighbors. Thus, all nodes have predictive and communicative IFP, while only actuated nodes have forward paths. In total, for a system with $n$ nodes and $m$ actuators, we will have $m$ forward paths, and $n$ predictive IFPs. Assume that node $i$ communicates with $n_i$ neighbors (this is determined by the sparsity of the relevant row of $R(k)$ and $M(k)$). Then, we will have $\sum_i n_i$ communicative IFPs. For our ring \eqref{eq:system_matrices} with $n=8$, $m=4$, $n_i=2$, we have 4 forward paths, 8 predictive IFPs, and 16 communicative IFPs. For this system, there are twice as many predictive IFPs as forward paths, and 5 times as many total IFPs as forward paths. The number of IFPs relative to the number of forward paths becomes even more drastic when we consider systems for which $m \ll n$, which are common in biology; organisms sense far more information than they are able to act upon. For example, a human can see objects that are hundreds of meters away, but can only act on objects within a small radius around his or her body -- and even then, in a manner severely limited by anatomy, mobility, and strength.

Thus, for highly underactuated systems, such as organisms, our SLS mesocircuit predicts a very large amount of IFPs relative to forward paths, consistent with experimental observations \cite{Budd1998}. IFPs in our mesocircuit carry signals predicting future sensor input; they are necessary at every sensor, as opposed to forward paths, which are only necessary for actuated nodes. Additionally, predictive signals in IFP use information of past \textit{actions}; this is consistent with experimental data showing motor-related activity in visual areas \cite{Stringer2019, Musall2019}. Overall, localized and distributed processing are the main features that motivate IFPs in our mesocircuit. These features are not captured in standard models involving IFPs. Although other models (e.g. modulation and memory processes, gain control, \cite{Muckli2013} predictive coding \cite{Huang2011}, recurrent neural networks \cite{Nayebi2018}, Bayesian estimation \cite{Paper2}) contain some IFPs, our model produces the largest amount of IFPs. In our model, IFPs are not merely supplements to the forward path; they are crucial for proper function of the mesocircuit.

We remark that this mesocircuit is based on the standard realization of the state feedback SLS controller, which is internally stable \cite{Anderson2019}. IFPs are a central and necessary feature in all known alternative SLS realizations \cite{Anderson2017, Tseng2020}, as well as full control and output feedback SLS controllers. We anticipate that for any local and distributed implementation of the SLS controller, IFPs will outnumber forward paths; we defer a more thorough exploration of implementation details to future work. In addition to forming the basis for this mesocircuit, SLS also enjoys unique scalability benefits, which will be useful for producing large-scale models.

	\section{Conclusion} \label{sec:conclusion}

From \textit{a priori} principles of local processing, local reaction, and underactuation, our SLS-based mesocircuit predicts a high ratio of internal feedback pathways (IFPs) to forward pathways. This has striking resemblance to neuroanatomy; our model complements data-driven techniques and provides human-interpretable insights on the function of IFPs in the nervous system. Future work will aim to use this mesocircuit to produce concrete, quantitative predictions for biology; potential experimental settings are discussed in \cite{Paper1}.
	
	\bibliography{internal_feedback}
	\bibliographystyle{IEEEtran}
\end{document}